\documentclass[%
reprint,
 amsmath,amssymb,
 prl,
]{revtex4-2}

\usepackage{graphicx}
\usepackage{dcolumn}
\usepackage{bm}
\usepackage[x11names]{xcolor}

\usepackage{pifont} 
\usepackage{makecell} 
\usepackage{etoolbox} 
\usepackage{lipsum} 
\usepackage[capitalize]{cleveref}
\usepackage{scrextend}
\usepackage{tikz}
\usepackage{pgfplots}
\usepackage{pdfpages}
\usepackage{pgffor}

\usepgfplotslibrary{external}
\pgfplotsset{compat=1.18}
\crefformat{footnote}{#2\footnotemark[#1]#3}

\makeatletter
\appto{\appendix}{%
  \@ifstar{\def\theequation@prefix{A.}}%
          {}%
}
\makeatother

\newcommand\sbullet[1][1.5]{$\mathbin{\vcenter{\hbox{\scalebox{#1}{$\bullet$}}}}$}
\newcommand{\yes}{\color{Turquoise4}\sbullet\color{black}}
\newcommand{\no}{\color{OrangeRed3}\sbullet\color{black}}
\newcommand{\maybe}{\color{Goldenrod1}\sbullet\color{black}}

\makeatletter
\AtBeginDocument{\let\LS@rot\@undefined}
\makeatother

\begin{document}


\title{Rare Event Sampling using Smooth Basin Classification}

\author{Sander Vandenhaute}
 \affiliation{Center for Molecular Modeling, Ghent University (BE)}
\author{Tom Braeckevelt}
 \affiliation{Center for Molecular Modeling, Ghent University (BE)}
\author{Pieter Dobbelaere}
 \affiliation{Center for Molecular Modeling, Ghent University (BE)}
\author{Massimo Bocus}
 \affiliation{Center for Molecular Modeling, Ghent University (BE)}
\author{Veronique Van Speybroeck}%
 \email{veronique.vanspeybroeck@ugent.be}
 \affiliation{Center for Molecular Modeling, Ghent University (BE)}

\date{\today}

\begin{abstract}
The efficiency of atomic simulations of materials and molecules can rapidly deteriorate when large free energy barriers exist between
local minima.
We propose smooth basin classification, a universal method to define reaction coordinates based on the internal feature representation of a graph neural network.
We achieve high data efficiency by exploiting their built-in symmetry and adopting a transfer learning strategy.
We benchmark our approach on challenging chemical and physical transformations, and show that it matches and even outperforms reaction coordinates defined based on human intuition.
\end{abstract}

\maketitle

Dynamic simulations of materials and molecules can provide both qualitative and quantitative information on chemical and physical transformations.
While explicit integration of Newton's equation of motion requires timesteps in the order of a femtosecond, the time required for many transformations to complete is often beyond the micro- or millisecond range
because of large energy barriers between initial and final state.
Such transformations are therefore almost always rare events, and their simulation can easily require billions of time steps \cite{frenkel2023understanding, tuckerman2010statistical}.

Over the past decades, a variety of advanced sampling techniques have been proposed in order to improve the sampling efficiency of dynamic simulations.
A particularly intuitive approach is to bias the dynamics along certain directions in phase space,
effectively lifting the system over obstructive free energy barriers.
The bias energy is defined along a certain predefined reaction coordinate or collective variable (CV).
The CV is a function of the atomic coordinates and should naturally discriminate between initial and final state without
violating the physical symmetries in the system \cite{peters2017}.
Specifically, the CV should remain invariant with respect to global translations and rotations, as well as permutations of atoms which are chemically identical.
In the vast majority of practical applications, such CVs are manually defined based on chemical and physical intuition, and they usually
contain simple invariants such as specific unit cell parameters or coordination numbers of one chemical species with respect to another.
However, manual definition can be highly challenging and time-consuming, because the metastable intermediate(s) and/or transition state(s) are not always known in
advance, or because the required invariances are difficult to satisfy \cite{Grifoni2019, Ahalawat2018, Niu2018}.

Considering the numerous achievements of machine learning in atomistic modeling,
it is sensible to anticipate that suitable reaction coordinates can also be learned from (unbiased) simulation data,
and ideally without requiring human intuition about the transition mechanism(s).
A wide variety of such methods have been proposed over the past decade, and we briefly mention a few notable examples
while referring the reader to specialized reviews for a more comprehensive discussion \cite{Chen2021, Bhakat2022, Tiwary2024}.
Supervised methods learn a continuous discriminant function which separates
free energy minima based on user-defined structural descriptors, for example using (deep) linear discriminant analysis \cite{Elishav2023, Bonati2020, Mendels2018, Ray2023} or (variational) autoencoders \cite{sipka2023, Zou2024, Wang2021}.
Unsupervised methods do not directly learn a discriminant function but instead identify so-called slow modes,
either within a given set of descriptors or entirely end-to-end;
see e.g. time-lagged independent component analysis or time-lagged autoencoders \cite{Hernandez2013, Wehmeyer2018, Mardt2018}.
While these methods demonstrate the potential of using machine learning for defining reaction coordinates, their widespread adoption has remained rather limited.
We hypothesize that this is due to inherent limitations regarding the required volume of simulation data, the difficulty of constructing an appropriate set of input features,
or the difficulty in interpreting and validating the obtained CVs.

\begin{figure*}[t!]
    \centering
    \includegraphics[scale=1.60]{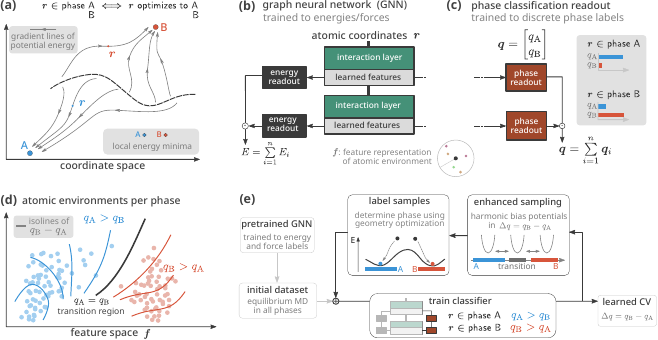}
    \caption{
    \textbf{(a)} Illustration of the partitioning of the phase space into basins of attraction. Each point $\bm{x}$ can be uniquely associated to its basin, i.e. its local potential energy minimum.
    \textbf{(b)} Base GNN which is pretrained to reference energies and forces.
    The interaction layers effectively generate a learned feature representation $\bm{f}$ for the atomic environments.
    \textbf{(c)} a phase classifier which is trained separately on atomic geometries which are labeled with the 
    associated basin as in (a).
    \textbf{(d)} Illustration of the feature space embedding of atomic environments based on their (local) phase, together with the isolines of $\Delta q$.
    \textbf{(e)} Active learning loop for SBC; the log probabilities $\bm{q}$ are used to bias
    the dynamics from one basin to another, in order to sample geometries along the transition path(s). The gathered data is labeled with the correct basin of attraction (a) after which the classifier is retrained. See the Appendix for more information.
    }
    \label{model}
\end{figure*}

In this work, we propose smooth basin classification (SBC); a universal method to construct CVs and compute free energy differences for chemical and physical transformations.
SBC builds upon the successful development of graph neural networks (GNNs) as effective interatomic potentials by
using their learned feature space as ansatz for constructing physically meaningful CVs \cite{Thomas2018, Batzner2022, Kovacs2023,tan2024enhanced}.
Our key finding is that GNN-based interaction potentials produce a feature representation for atomic environments which
can be used to drive transitions between free energy minima and characterize their relative stability.

Consider a system with two basins of attraction, named A and B.
Each point $\bm{r}$ in the system's coordinate space can be uniquely labeled with A or B depending
on the trajectory of a geometry optimization; $\bm{r}$ belongs to basin A whenever it optimizes to any of the local minima
within A, and vice-versa for B (Figure \ref{model}, a).
In SBC, we aim to construct a mapping between each configuration $\bm{r}$ and its associated basin of attraction.
Instead of defining this mapping directly in terms of the atomic coordinates -- which are not invariant with respect to the symmetry operations in the system --
we embed it into the learned feature space of a GNN which has been pretrained to reproduce the potential energy surface of the system (Figure \ref{model}, b).
In general, the interaction layers of the GNN parse the chemical environment of atom $i$ into a feature vector $\bm{f}_i$ in this space.
Importantly, this feature vector is rigorously invariant with respect to all symmetry operations in the system, and is specifically trained to describe the atomic interactions.
Both the per-atom energy $E_i$ and the total energy $E$ are computed
from the feature vectors $\bm{f}_i$ using a simple readout function:
\begin{align}
\bm{r}\quad \xrightarrow[]{\text{GNN}}\quad \{\bm{f}_i\}_{i=1}^n\quad  \xrightarrow[]{\text{energy readout}} \quad E = \sum_{i=1}^n E_i
\label{energy_readout}
\end{align}
in which $n$ denotes the total number of atoms in configuration $\bm{r}$.
Because the features $\bm{f}_i$ are maximally informative for determining the interaction energy of a given atomic environment,
we may assume that they are also highly informative in predicting the basin of attraction.
In full analogy with equation \ref{energy_readout}, we can augment
the GNN with an additional classification readout function (Figure \ref{model}, c):
\begin{align}
\bm{r}\quad \xrightarrow[]{\text{GNN}}\quad \{\bm{f}_i\}_{i=1}^n\quad  \xrightarrow[]{\text{phase readout}} \quad \bm{q} = \sum_{i=1}^n \bm{q}_i
\label{phase_readout}
\end{align}
in which we introduce the per-atom log probabilities $\bm{q}_i$ as an invariant tuple which measures how strongly the model
associates a given atomic environment $\bm{f}_i$ to basin A or B -- as commonly done in a classification setting \cite{Bishop2006}:
\begin{align}
\bm{q}_i = \Big(q_i^A, q_i^B\Big)
\end{align}

\begin{figure*}[t]
    \centering
    \includegraphics[scale=0.93]{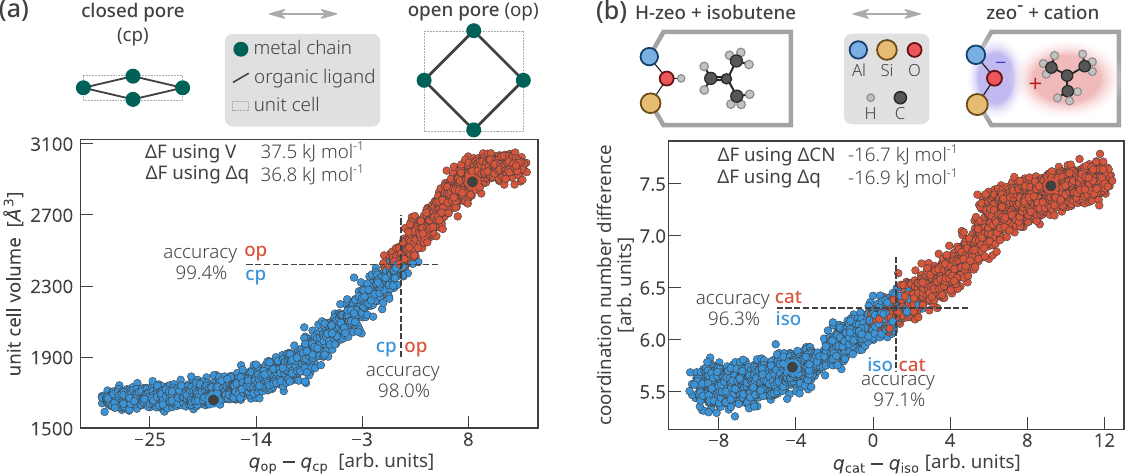}
    \caption{\textbf{(a)} The nonreactive phase transition in  MIL-53(Al).
    The scatter plot shows the distribution of both the unit cell volume $V$ as well as the log probability difference
    $\Delta q = q_\text{\sffamily op} - q_\text{\sffamily cp}$.
    Each dot represents a single atomic geometry, and its color represents whether it minimizes to the $op$ or $cp$ phase.
    Black dots indicate potential energy minima.
    \textbf{(b)} A proton transfer reaction in a zeolite. The scatter plot shows the distribution of both the coordination number difference as well as
    the log probability difference $\Delta q = q_{\text{\sffamily cat}} - q_\text{\sffamily iso}$, whereby the color again indicates the corresponding basin of attraction.
    The reported relative free energy differences are computed based on the free energy profiles in the Appendix.
   } 
    \label{scatters}
\end{figure*}

For example, if the model believes that a given atomic environment $i$ is more likely to appear in phase $A$ as opposed to phase $B$,
it will predict log probabilities $\bm{q}_i$ such that $q_i^A > q_i^B$.
After summing the per-atom log probabilities $\bm{q}_i$ to obtain the total log probabilities $\bm{q}=\left(q^A, q^B\right)$, 
the predicted basin of attraction of the total configuration is then determined by 
the largest component in $\bm{q}$.
A detailed overview of the phase classification readout can be found in the Appendix.
%

While the prediction of the basin of attraction is essentially a discrete classification task, the actual log probabilities $\bm{q}_i$ are
smooth functions of the atomic feature vectors $\bm{f}_i$ and hence the atomic coordinates --
which means that they can be used to introduce biasing forces (and biasing stress).
Ideally, we would like to design this bias such that it can drive transitions from and to any of the basins.
To this end, consider the difference between two log probabilities:
\begin{align}
    \Delta q &= q^B - q^A
    \label{cv}  
\end{align}

In essence, $\Delta q$ measures how much more likely the model thinks a given configuration belongs to basin B rather than basin A.
It is therefore strictly positive in basin B and strictly negative in basin A provided that the phase readout is properly trained (Figure \ref{model}, d).
Clearly, the quantity $\Delta q$ correctly discriminates between two basins of attraction while also being continuously differentiable with respect to the atomic coordinates (using standard backpropagation).
It is therefore, by definition, a valid collective variable for the transition between basins A and B.

In what follows, we discuss the application of SBC for prototypical physical and chemical transformations.
We use MACE as base GNN \cite{batatia2022,Kovacs2023}, and employ active learning as in Figure \ref{model} (e) to train the classifier
readout -- see the appendix for additional computational details.

Figure \ref{scatters} (a) shows a nonreactive phase transition in a solid-state system named MIL-53(Al), a flexible metal-organic framework based on a winerack topology \cite{Serre2002}. 
The system exhibits two basins of attraction which are relevant for tuning the material towards gas storage and/or sensing
applications; a closed pore ($cp$) and open pore ($op$) phase \cite{Vanduyfhuys2018}.
In these types of transitions, i.e. those for which the unit cell parameters vary significantly between the different basins, researchers often employ some combination of unit cell parameters as collective variable -- in this case, the unit cell volume \cite{Demuynck2018}.
Figure \ref{scatters} (a) demonstrates the performance of the unit cell volume as formal discriminator between the two basins, and
compares it with a difference of learned log probabilities as proposed in equation \ref{cv}.
The misclassified geometries are all localized
near the transition region (such they will not cause substantial differences in the relative stability).
Note that the network learns to distinguish low- and high-volume configurations based on atomic environments only, i.e. without explicitly relying on unit cell parameters as input.

\begin{table*}[!t]
\def\arraystretch{1.0}
\begin{ruledtabular}
\begin{tabular}{lccccccc}
&
\multicolumn{2}{c}{training cost} &
\makecell{input \\ descriptors} &
\multicolumn{3}{c}{symmetries} &
\makecell{initialized without \\ prior transition path(s)} \\
& snapshots & time [ns] & & trans & rot & perm & \\
\colrule
time-lagged AE \cite{Wehmeyer2018} & & 750 ns &
learned & \maybe & \maybe & \no & \\
FEBILAE \cite{Belkacemi2022} & & 1500 ns & learned &
\maybe
& \maybe & \no & \\
VAMPnets \cite{Mardt2018} & 250,000 & 250 ns & learned &  \maybe & \maybe & \no & \ding{51} \\
LINES \cite{Odstrcil2022} & & 100 ns & manual &  \yes & \yes & \no & \\
LED \cite{Vlachas2022} & & 100 ns & manual & \yes & \yes & \no & \\
Ref. \cite{Sun2022} & $>$300,000 & & learned & \maybe & \maybe & \no & \ding{51} \\
Deep-LDA \cite{Bonati2020} & 10,000 & & manual & \yes & \yes & \no & \\
Deep-TDA \cite{Trizio2021} &  & 40 ns & learned& \yes & \yes & \no & \ding{51} \\
Ref. \cite{Bonati2021} & 15,000 & 15 ns & manual & \yes & \yes & \no & \ding{51} \\
Ref. \cite{mullender2023effective} & 8,000 & 512 ns & manual & \yes & \yes & \no & \\
\colrule
SBC & \textbf{3,694} & \textbf{11 ns} & learned & \yes & \yes & \yes & \ding{51} \\
\end{tabular}
\end{ruledtabular}
\caption{Performance of various CV learning methods on alanine dipeptide. Key characteristics are the required amount of input data
(lower is better), the built-in symmetry as inductive bias, and whether or not an initial
transition path is required for the model to learn the transition. Missing values indicate that the corresponding quantity
was not directly reported in either the main manuscript or any of its appendices.
Yellow dots indicate that the required invariance was achieved by means of 
a manual structural alignment procedure -- see the corresponding references for more information.}
\label{comparison}
\end{table*}

For global phase transformations in solids such as the one in Figure \ref{scatters} (a), a size-extensive definition of $\bm{q}$ as a 
sum over per-atom log probabilities $\bm{q}_i$ is intuitively sensible because it gives the model the ability to assign nonzero log probability
contributions to all atoms simultaneously.
However, in chemical transformations, the rare event is most often a local process whose free energy barrier
does not necessarily depend on the size of the surrounding environment (e.g. the catalyst).
To demonstrate the applicability of our approach to those cases, we investigate a proton
transfer reaction from a Br{\o}nsted acid site in a zeolite to an alkene guest (Figure \ref{scatters}, b). 
The reference collective variable in this case is a coordination number difference which is defined explicitly in terms of
the indices of the oxygen atoms of the active site as well as the indices of the carbon atoms of the molecule and all of the hydrogens -- see the Supplemental Material (Section S1.2) for more details.
Classification accuracies are excellent for both the proposed CV based on chemical intuition as well as
the learned log probability difference.
Misclassified geometries are localized at or near the transition region, which means that they will not induce discrepancies in the relative stability.
Indeed, for both systems, free energy calculations based on either the reference or the learned CV yield the same relative stability (to within 1 kJ/mol).

\begin{figure}[t]
    \centering
    \includegraphics[scale=1.1]{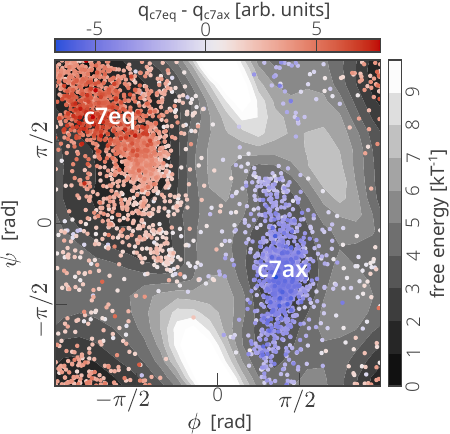}
    \caption{Free energy landscape of alanine dipeptide as obtained from umbrella sampling
    in the conventional backbone dihedral angles $\phi$ and $\psi)$.
    The scattered dots indicate the training data distribution as obtained during active learning of the SBC model; the color-coding
    indicates the $\Delta q$ values after final training.
    See the Supplemental Material (Section S1) for additional computational details.
    }
    \label{alanine}
\end{figure}

Note that the quantity $\Delta q$ satisfies the same symmetries as
the potential energy $E$ of the system (Figure \ref{model}, b/c).
As such, the learned collective variable is invariant with respect to
global translations and rotations as well as permutations of chemically identical atoms.
Achieving permutational symmetry in particular is quite remarkable, and should be contrasted with
both common practice CVs and most other ML-based approaches \cite{Bonati2020, Mardt2018, Wehmeyer2018, Trizio2021}
which often use functions and/or input features that depend explicitly on the atomic indices
(thereby violating permutation invariance)
or specific components of the unit cell vectors (thereby violating rotational as well as periodic invariance).
We hypothesize that the Euclidean and permutation symmetry of $\Delta q$ is the primary reason for the exceptional data efficiency
of the method, alongside the effective active learning scheme (Figure \ref{model}, e) and the implicit use of equivariant features within the GNN.
We demonstrate this explicitly based on a case study on alanine dipeptide, which represents a topical nonreactive conformational
change that has served as enhanced sampling benchmark system in many
studies \cite{Wehmeyer2018,Belkacemi2022,Mardt2018,Odstrcil2022,Vlachas2022,Sun2022,Bonati2020,Bonati2021,mullender2023effective,kang2024computing}.
Table \ref{comparison} enumerates and compares the most prominent CV learning methods with SBC
for alanine dipeptide.
We evaluate a number of key characteristics,
including the required amount of input data to train
a given CV, its transferability towards symmetrically equivalent geometries, whether or not it requires manual feature selection,
and whether or not an initial transition path is required.
In essence, Table \ref{comparison} shows that SBC outperforms every single method in all of these aspects.
It requires the fewest structures and smallest total simulation time; it is the only method that is rigorously invariant
with respect to all symmetry operations; and it does not require any kind of manual input feature selection or initial transition path.
Figure \ref{alanine} demonstrates how SBC is capable of discovering several transition paths during its active learning loop.
In Section S1.3 of the Supplemental Material,
we further show that the relative free energy difference as computed using $(\phi, \psi)$ and $\Delta q$
agrees within 1 kJ/mol.


We conclude with a discussion on the current limitations of SBC as well as directions for future research.
Clearly, the success of our method is critically dependent on the availability of ML potentials which can predict the atomic interactions in the system with high fidelity.
In this work, we considered both self-developed and foundation models as base GNN in which the SBC model is inserted --
the alanine dipeptide and MIL-53(Al) case studies were performed with MACE models trained using psiflow \cite{Kovacs2023,Vandenhaute2023},
while a MACE foundation model was used for the proton transfer reaction\cite{batatia2023foundation}.
As such, we can conclude that SBC is fundamentally compatible with both.
However, more challenging scenarios might require
multi-task training of the entire GNN (including its interaction layers), in which each sample in the training set is labeled with QM reference data and additionally
a phase or basin label -- this will incentivize the model to design the per-atom feature embedding to be informative for both
potential energy prediction as well as basin classification.
In addition to relative free energies, the learned CVs can be employed to predict the kinetics of rare events based on e.g.
reactive flux and/or path sampling methods \cite{peters2017_}.
Finally, we mention that the proposed method can be extended towards the more general case of $n$ free energy minima
simply by increasing the number of classes in the classification layer.








\begin{center}
\textbf{\small DATA AVAILABILITY STATEMENT}
\end{center}
An implementation of SBC in MACE will be made available in a public GitHub repository upon final publication; for peer-review purposes, we also provide the code through the following link: \url{https://tinyurl.com/mufvj4ur}.
All additional input files, all generated datasets, and all of the models which were trained in this manuscript will be made available in a public Zenodo archive upon final publication; for peer-review purposes, these are available upon request as well. \\

\begin{center}
\textbf{\small ACKNOWLEDGEMENTS}
\end{center}
The authors acknowledge funding from the Research Board of Ghent University (V.V.S) and iBOF-21-085 PERSIST (V.V.S. and T.B.).
S.V. and P.D. wish to thank the Research Foundation – Flanders (FWO) for doctoral fellowships (grant nos. 11H6821N and 11O2123N respectively).
M.B. acknowledges financial support from the Fund for Scientific Research Flanders and the Excellence of Science (EOS) Project BioFact (EOS ID 30902231).
We acknowledge the EuroHPC Joint Undertaking for awarding this project access to the EuroHPC supercomputer LUMI, hosted by CSC (Finland)
and the LUMI consortium through a EuroHPC Regular Access call.
Part of the simulations were performed on the Luxembourg national supercomputer MeluXina.
The authors gratefully acknowledge the LuxProvide teams for their expert support.


\appendix

\section{Appendix on the Model Architecture}
In analogy with the prediction of the total potential energy $E$ as a sum of predicted atomic energies $E_i$,
we predict the total log probability $\bm{q}$ as a sum of per-atom log probabilities $\bm{q}_i$
(see Equation 2 in the main text).
In a system with $N$ basins, the log probabilities $\bm{q}_i$ are a vector of length $N$, i.e. there is one per-atom logit for each basin.
We use $\bm{f}_i^{(j)}$ to denote the GNN features from layer $j$ for atom $i$.
First, we forward the node features into a MLP per layer:
\begin{align}
    \bm{h}_i^{(j)} = \text{\sffamily MLP}^{(j)}\left(\bm{f}_i^{(j)}\right)
\end{align}
Next, we concatenate $\bm{h}_i^{(j)}$ for all GNN layers $j$,
and use a final linear layer with weights $W$ to compute the
per-atom log probabilities $\bm{q}_i$:
\begin{align}
    \bm{q}_i = W\left( \oplus_j \bm{h}_i^{(j)}\right) 
    \label{energycorrection}
\end{align}
Finally, the total log probabilities $\bm{q}$ are obtained by summing the individual $\bm{q}_i$
\begin{align}
\bm{q} = \sum_{i=1}^n \bm{q}_i
\end{align}
The phase readout layers are analogous to the existing energy readouts found in conventional GNNs \cite{Thomas2018, Batzner2022, Kovacs2023}
(Figure \ref{model}, c),
and essentially consist of a multilayer perceptron (MLP) containing at most two nonlinearities and about $\sim$1000 weights in total
(compared to $\sim$100,000 weights in the GNN).

\section{Appendix on Model Training}
The predicted log probabilities $\bm{q}$ are normalized and trained to atomic geometries which are labeled with the correct basin.
The label is obtained by minimizing the energy of the structure, as explained in Figure 1 in the main manuscript.
In a classification setting, one often uses the cross-entropy loss function \cite{Bishop2006}.
We can define the phase label $\bm{p}$ of an atomic geometry as a one-hot vector which has the same length as
$\bm{q}$ (i.e. the number of basins), and whose components indicate the correct basin of attraction:
\begin{align}
p^{\mathcal{C}} = \begin{cases}
    1 \quad \text{if }\text{phase} = \mathcal{C} \\
    0 \quad \text{otherwise}
\end{cases}
\label{phaselabel}
\end{align}
Based on Equation \ref{phaselabel}, we can write the cross-entropy loss as a simple scalar product:
\begin{align}
    \text{CE}(\bm{q}) = - \bm{p}\cdot\bm{q}
\end{align}
The classifier is trained while keeping the GNN and energy readouts fixed, similar to conventional transfer learning approaches.
Its training data consists of atomic geometries which are labeled with their corresponding basin of attraction (e.g. A or B).

Note that the cross-entropy loss is a function defined on the total log probabilities $\bm{q}$, 
not on the specific per-atom log probabilities $\bm{q}_i$.
This implies that the cross-entropy loss will incentivize the model to predict correct total log probabilities, but otherwise introduces
no restrictions regarding the specific per-atom log probabilities $\bm{q}_i$.
This makes the model prone to overfitting, especially in the low-data regime.
To improve this, we considered a number of regularization strategies
including label smoothing, an L2 penalty on the weights, and an L1 penalty on logit gradients with respect to the atomic coordinates.
Their efficacy was highly dependent on both the system under study and their specific weights in the loss function,
and we chose not to apply them.
Instead, we designed a new regularization term which prevents the per-atom log probabilities $\bm{q}_i$
from becoming strongly positive for any given atomic environment $i$.

First, note that the global probability for each phase can be computed based on the predicted
log probabilities $\bm{q}$ using a softmax function:
\begin{align}
    P\left[\text{A}\right] &= \frac{\displaystyle e^{q^\text{A}}}{\displaystyle \sum_\mathcal{C} e^{q^\mathcal{C}}}
    \label{classprobability}
\end{align}
for any class A.
We can apply a similar reasoning to the per-atom log probabilities $\bm{q}_i$ in order to obtain per-atom phase probabilities $P_i[A]$.
Next, consider
\begin{align}
    \mathbb{E}[\bm{q}_i] = \sum_\mathcal{C} P_i[\mathcal{C}] q_i^\mathcal{C}
    \label{expectation}
\end{align}
The quantity $\mathbb{E}[\bm{q}_i]$ denotes an expected per-atom logit value.
In practice, given the exponential dependence between the probabilities and
the log probabilities, there will only be one nonzero contribution in the sum in Equation \ref{expectation}, and the
expectation value will essentially reduce to the largest (per-atom) logit value.
Note that the predicted per-atom phase (based on the per-atom log probabilities)
can differ from the predicted global phase (based on the total log probabilities). 
Finally, in the loss function, we add an L2 regularization on $\mathbb{E}[\bm{q}_i]$, which essentially
prevents the model from being overconfident in any of the per-atom log probabilities:
\begin{align}
\mathcal{L} = CE(\bm{q}) + \lambda \sum_{i=1}^n ||\mathbb{E}[\bm{q}_i]||_2
\end{align}
For all experiments in this work, we use $\lambda = 1$.

Furthermore, note that the actual training of the phase readouts can be made very efficient
because the GNN node features $\bm{f}_i$ can be cached and reused at every step of the optimization
since the weights of the pretrained GNN are fixed.
As a result, training is very efficient, and all SBC models in this work can be trained within one hour on a consumer-level GPU.

\begin{figure*}[t]
    \centering
    \includegraphics[scale=0.93]{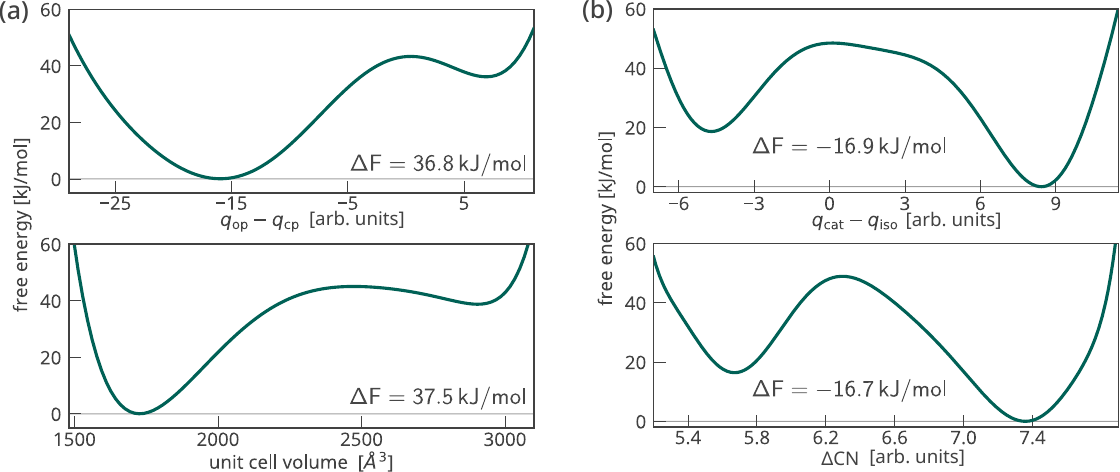}
    \caption{Helmholtz free energy profiles for \textbf{(a)} the phase transition in  MIL-53(Al) and
    \textbf{(b)} the proton transfer reaction, as obtained from umbrella sampling in either the learned collective variable $\Delta q$ or the reference collective variable.} 
    \label{FES}
\end{figure*}

\section{Appendix on the Active Learning}
\label{appendixlearning}
Figure 1 (e) in the main manuscript outlines the active learning workflow in more detail.
We start by training the phase readouts on equilibrium samples from each of the free energy minima, after which we begin with the active learning loop and perform enhanced sampling.
We introduce harmonic restraints centered on different values of $\Delta q$ in order to force the system to transition from one basin
to another.
In each iteration, we position 32 umbrellas along the entire CV range, at three different temperatures as to ensure that both low- and (relevant) high-energy regions are properly sampled.
In each umbrella, the system is randomly initialized in either of the two minima.
The samples from the resulting trajectory are labeled according to their basin of attraction,
as determined by a simple energy minimization for each sample.
After this, the model is retrained on all data, and the procedure is repeated.
See the Supplemental Material (Section S1) for a complete overview of all hyperparameters.

\section{Appendix on the Graph Neural Network}
\label{appendixdetails}
All calculations in this work employed MACE as base potential \cite{Kovacs2023}, although we should emphasize that SBC is independent of the specific GNN architecture \cite{duval2023hitchhikers}.
The interatomic potentials for MIL-53(Al) and alanine dipeptide were trained from scratch using psiflow, at the PBE-D3(BJ) level of theory and with a triple zeta basis set \cite{Vandenhaute2023}.
DFT evaluations were performed with CP2K 2023.1 for MIL-53(Al) in which the basis set was augmented with plane waves up until a cutoff of 1000 Ry, and ORCA 5.0.4 for alanine dipeptide \cite{Kuhne2020, Neese2020}.
For the proton transfer reaction in the zeolite catalyst, we employed a foundational MACE model \cite{batatia2023foundation}.
For the MACE models, we employ node features with $\ell_\text{max} = 2$ and 16 channels each, with a cutoff radius of 6.5 \AA{}.
All models will be made available upon final publication.

\section{Appendix on the Free Energy Calculations}
Using the log probability difference $\Delta q = q_\text{\sffamily op} - q_\text{\sffamily cp}$ as CV (see equation \ref{cv}), we can employ
conventional enhanced sampling techniques and evaluate the free energy profile $F(\Delta q)$.
Here, we choose to employ umbrella sampling with harmonic restraints on the CV:
\begin{align}
    U_\text{\sffamily bias}(\bm{r}, \bm{h}) = \frac{K}{2} \Big[\Delta q\left(\bm{r}, \bm{h}\right) - q_0\Big]^2
    \label{umbrella}
\end{align}
in which we emphasize that the bias energy contains an implicit dependence on both the atomic coordinates $\bm{r}$ and the unit cell parameters $\bm{h}$ through the learned atomic environment features $\bm{f}_i$ \cite{tan2024enhanced}.
We use the multistate Bennett acceptance ratio (MBAR) equations to combine samples from multiple umbrellas in a statistically optimal manner and
obtain free energy profiles as shown in Figure \ref{FES} \cite{Shirts2008, shirts2017reweighting, Shirts2020, pymbar}.
Note that while relative stabilities are intrinsic to the system, the actual free energy profiles depend on the particular choice of CV, so the learned and reference free energy profiles do not need to have the same shape (although there exist ways to transform one profile into another \cite{Bailleul2020}).

\bibliography{apssamp}



\section*{Supplementary material}

# Free Energy Calculations using Smooth Basin Classification: Supplemental Material

Sander Vandenhaute[1], Tom Braeckevelt[1], Pieter Dobbelaere[1], Massimo Bocus[1], and Veronique Van Speybroeck[1]

[1]Center for Molecular Modeling, Ghent University (BE)

May 24, 2024

## S1 Additional Computational Details

All molecular dynamics simulations were performed using Langevin temperature and/or pressure control. Bias potentials on traditional CVs were added via PLUMED [1]; bias potentials on $\Delta q$ were added manually using a modified MACE-ASE interface. Table S1 provides an overview with the most important hyperparameters for each case study. Note that there is little system-specific tuning necessary – all systems use three different temperatures, similar sampling times, similar bias strengths, and identical SBC parameters.

### S1.1 MIL-53(Al)

MIL-53(Al) is a metal-organic framework which exhibits a closed pore ($cp$) and large pore ($lp$) phase (Figure S1). In previous work, we showed that the $op$-to-$cp$ transition can proceed spontaneously at room temperature conditions and pressures above approximately 30 MPa [3]. Since both phases are characterized by a different unit cell shape, the transition between them necessarily requires anisotropic unit cell dynamics during active learning. As a result, the geometry optimizations which are used to label each of the samples with either a $cp$ or an $op$ label relax both atomic coordinates as well as unit cell components. In spite of the (small) unit cell shape change during the transition, the unit cell volume is a good collective variable for this transition, as can be seen in Figure 2 in the main manuscript; the volume appears to be a near-perfect discriminator for predicting the basin of attraction. The reference free energy profile was computed with umbrella sampling in the unit cell volume; we applied harmonic bias potentials centered at 64 equidistant volume points between $V = 1450$ Å$^3$ and $V = 3300$ Å$^3$, and with a force constant of 0.0045 kJ/(mol· Å$^6$).

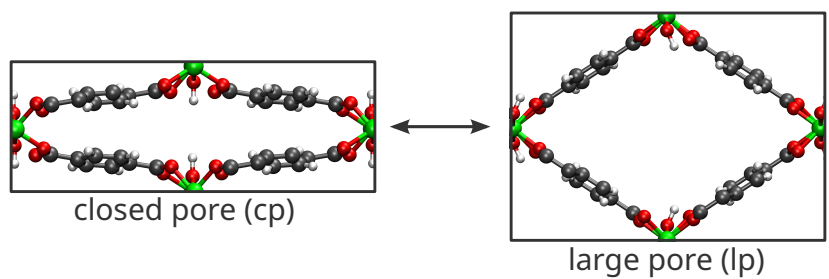

Figure S1: Atomic geometry of the closed pore (cp) and large pore (lp) of MIL-53(Al), a flexible metal-organic framework with the winerack topology [2]. It consists of an aluminum hydroxide chain which is interconnected with terephthalate organic linkers.

1

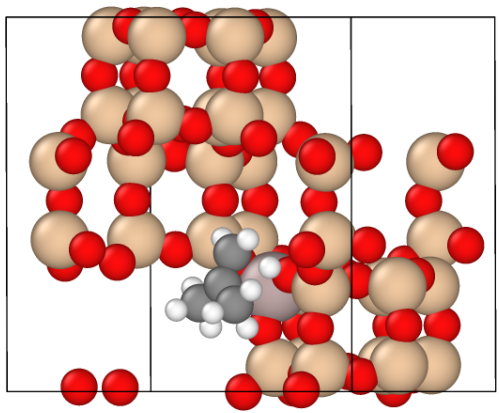

Figure S2 : Unit cell of chabazite with an aluminum active site. A proton transfer reaction can occur from the framework to an isobutene guest molecule inside the largest cage.

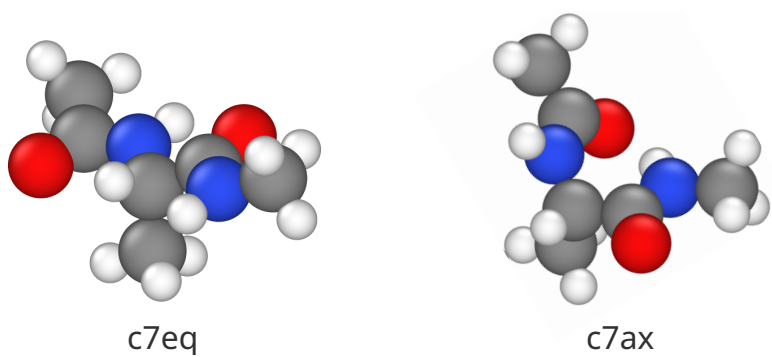

Figure S3 : Atomic geometry of alanine dipeptide.

## S1.2  isobutene @ CHA

We consider an isobutene guest in the largest cage of chabazite, a relatively small zeolite framework (Figure S2). The active site of the reaction (technically a Brønsted acid site) is created by replacing a silicon atom with an aluminum atom, and adding a proton to one of its neighboring oxygens. The reference collective variable as mentioned in Figure 2 (b) in the main manuscript is a simple coordination number difference. With $r_0 = 1.4$ Å and $A$, $B$ specific groups of atoms, we have:

$$CN(A, B) = \sum_{\substack{i \in A \\ j \in B}} \frac{1 - \left(\frac{r_{ij}}{r_0}\right)^6}{1 - \left(\frac{r_{ij}}{r_0}\right)^{12}} \quad (1)$$

The coordination number difference is then defined between the coordination of the three outer isobutene carbons and all hydrogens, and the coordination of the four active site oxygens and all hydrogens. The reference free energy profile was computed with umbrella sampling in the coordination number difference; we applied harmonic bias potentials centered at 64 equidistant points between $CV = 5.2$ and $CV = 7.9$, and with a force constant of 1500 kJ/mol.

| | model | | ensemble | temperatures | bias potentials | | sampling times [ps] | | |
| --- | --- | --- | --- | --- | --- | --- | --- | --- | --- |
| | GNN | SBC layer sizes | | | # | kappa [eV$^{-1}$] | 1 | 2 | 3 |
| **MIL-53(Al)** 152 atoms, triclinic | system-specific | (16, 8) | **NPT** (P = 0 MPa) | 150 K 300 K 500 K | 32 | 0.1 | 96 | 384 | 384 |
| **isobutene @ CHA** 121 atoms, triclinic | foundation | (16, 8) | **NVT** | 150 K 300 K 600 K | 32 | 0.1 | 192 | 480 | 960 |
| **alanine dipeptide** 22 atoms | system-specific | (16, 8) | **NVT** | 300 K 600 K 900 K | 32 | 0.1/0.2 | 192 | 480 | 960 |

Table S1 : Overview of the SBC and active learning hyperparameters for each of the case studies performed in this work. SBC requires no tuning (the layer sizes and nonlinearities are identical for each of the three cases, even if the underlying GNNs were different), whereas the sampling requires some fine-tuning related to the temperature(s) and the sampling times for each of the three iterations.

### S1.3   alanine dipeptide

The two basins of attraction (referred to in the main text as `c7eq` and `c7ax`) are shown in Figure S3 . The reference free energy surface as a function of $\phi$ and $\psi$ was computed using 2D umbrella sampling. We used an extensive 24×24 grid of umbrella centers, with a force constant of 400 kJ/(mol·rad$^2$). The relative free energies were obtained by integrating over the two minima on the free energy surface, which yielded a value of -4.8 kJ/mol (indicating that `c7eq` is more stable than `c7ax`). Based on the learned log probabilities, we performed 1D umbrella sampling, using 64 umbrellas and with a force constant of about 20 kJ/mol. After integrating over the minima, we obtained a relative free energy difference of -4.5 kJ/mol, which is in excellent agreement with the reference value.



\end{document}